\documentclass[10pt,a4paper,dvipdfmx]{article}

\usepackage[dvipdfmx]{graphicx}

\usepackage{fancyhdr}
\usepackage{booktabs}
\usepackage{amsmath}
\usepackage{amssymb}
\usepackage{lscape}
\usepackage{bm}
\usepackage{float}
\usepackage{comment}
\usepackage{url}
\usepackage{hyperref}
\usepackage{calc}
\usepackage[top=30truemm,bottom=30truemm,left=28truemm,right=28truemm]{geometry}
\usepackage{setspace}
\usepackage{cite}
\usepackage{upgreek}
\usepackage{indentfirst}
\usepackage{multirow}
\usepackage{color}

\usepackage{mathrsfs}

\usepackage{caption}
\usepackage{titlesec}
\titleformat*{\section}{\normalsize\bfseries}
\titleformat*{\subsection}{\normalsize\itshape}

\let\OLDthebibliography\thebibliography
\renewcommand\thebibliography[1]{
  \OLDthebibliography{#1}
  \setlength{\parskip}{0pt}
  \setlength{\itemsep}{3pt plus 0.3ex}
}

\begin{document}

$\,$

\vspace{60pt}

\begin{spacing}{1.8}
\begin{flushleft}
{\bf\LARGE New Estimation of the Curvature Effect for the X-ray Vacuum Diffraction Induced by an Intense Laser Field}
\end{flushleft}
\end{spacing}

\vspace{-10pt}

\begin{flushright}
\begin{minipage}{0.85\hsize}
\begin{spacing}{1.2}
{\bf Y. Seino$^{\,1,}$*, T. Inada$^{\,2,}$, T. Yamazaki$^{\,3}$, T. Namba$^{\,2}$ and S. Asai$^{\,1}$}
\end{spacing}

\vspace{5pt}

{\small
$^{1}$ \quad Department of Physics, Graduate School of Science,

$\, \:$ \quad The University of Tokyo, 7-3-1 Hongo, Bunkyo-ku, Tokyo 113-0033, Japan\\
$^{2}$ \quad International Center for Elementary Particle Physics,

$\, \:$ \quad The University of Tokyo, 7-3-1 Hongo, Bunkyo-ku, Tokyo 113-0033, Japan\\
$^{3}$ \quad High Energy Accelerator Research Organization,

$\, \:$ \quad KEK, 2-4 Shirane Shirakata, Tokai-mura, Naka-gun, Ibaraki, 319-1195, Japan
}

\vspace{5pt}

E-mail: {\tt yseino@icepp.s.u-tokyo.ac.jp, tinada@icepp.s.u-tokyo.ac.jp}

\vspace{15pt}

{\bf Abstract:}
Quantum electrodynamics predicts x-ray diffractions under a high-intensity laser field via virtual charged particles, and this phenomenon is called as vacuum diffraction (VD).
In this paper, we derive a new formula to describe VD in a head-on collision geometry of an XFEL pulse and a laser pulse.
A wavefront curvature of the XFEL pulse is newly considered in this formula.
With this formula, we also discuss the curvature effect on VD signals based on realistic parameters at SACLA XFEL facility.

\end{minipage}
\end{flushright}

\vspace{8pt}

\hrule

\vspace{12pt}

\section{Introduction}\label{section1}
Photon-photon scattering is a nonlinear interaction between photons, intermediated by a virtual electron loop at the lowest order of quantum electrodynamics (QED).
To observe this process in the real vacuum, where the Coulomb fields of pre-existing charges do not contribute, many experimental approaches have been considered and carried out to excite this virtual loop.
A traditional way to stimulate the vacuum over a macroscopic scale, $\sim 1$ m, is to use permanent or electro- magnets that ``magnetize'' the vacuum \cite{c_PVLAS2016, c_BMV2014, c_OVAL2017}.
Another way, which does not apply such macroscopic-scale external fields, is to use a high-intensity laser that gives a very strong electromagnetic field to ``pump'' the vacuum \cite{c_MoulinGammaGamma1996, c_BernardFWM2000, c_BlinneVP2019}.
The pump laser locally and anisotropically changes the refractive index of the vacuum at its focus.
In order to detect this change, another probe light is collided against the focused pump laser.
The probe light is diffracted by this small focal spot, and its polarization is also changed.
These phenomena are called as ``vacuum diffraction (VD)'' and ``vacuum birefringence (VB)''.
X rays are suitable as the probe lights to detect the small vacuum structure because those have a shorter wavelength than optical lights.

VD and VB signals in various situations, e.g., different pump laser setups and collision geometries, have been calculated \cite{c_PiazzaVD2006,c_HeinzlVB2006,c_KryuchkyanVD2011,c_KingVD2010,c_DinuVP2014,c_KarbsteinVD2016,c_KarbsteinVD2018,c_SchlenvoigtVB2016, c_ShenVB2018}.
Because x-ray free-electron laser (XFEL) facilities with high-power laser stations \cite{c_SACLAStatus2017, c_Web_HiBEF, c_ShanghaiFEL2017} are appropriate to perform VD and VB experiments, some calculations \cite{c_SchlenvoigtVB2016, c_ShenVB2018,c_PiazzaVD2006,c_HeinzlVB2006, c_KarbsteinVD2016, c_KarbsteinVD2018} are specialized to experiments at these facilities.
However, these calculations do not include an angular divergence of a focused XFEL pulse, which is related to the curvature of the XFEL wavefront.
The beam divergence is typically ${\cal O}(10$---$1000)\mu$rad but is approximately equal to a signal divergence of VD.
Consideration of the curvature effect is important to predict the signal distributions in realistic setups.

In this paper, we derive a new formula for VD and VB signals in which the XFEL curvature effect is first taken into account.
The assumed setup for the calculation is a realistic condition in XFEL facilities, i.e., a focused pump laser pulse and a focused XFEL pulse collide in a head-on geometry.
Misalignments between both pulses are also included.

In Sec.~\ref{section2}, before considering the curvature effect, we derive a new simplified formula of VD and VB signals for XFEL experiments.
This approximated formula is useful to understand the parameter dependencies of VD and VB signals.
This new formula is validated in Sec.~\ref{section2.3} by comparing with the previous calculation \cite{c_KarbsteinVD2016} with real setups parameters at SACLA XFEL facility \cite{c_SACLAStatus2017}.
In Sec.~\ref{section3}, we calculate the curvature effect with this new formula.
The effect is estimated as the convolution of the XFEL angular distribution and the signal one.
Impacts of the curvature effect on the experiments are also discussed in Sec.~\ref{section3}.

\section{Simple formula of VD signal}\label{section2}

\subsection{Assumed setup for the calculation}\label{section2.1}

\begin{figure*}
  \begin{minipage}[b]{0.495\linewidth}
    \includegraphics[width = 1 \linewidth]{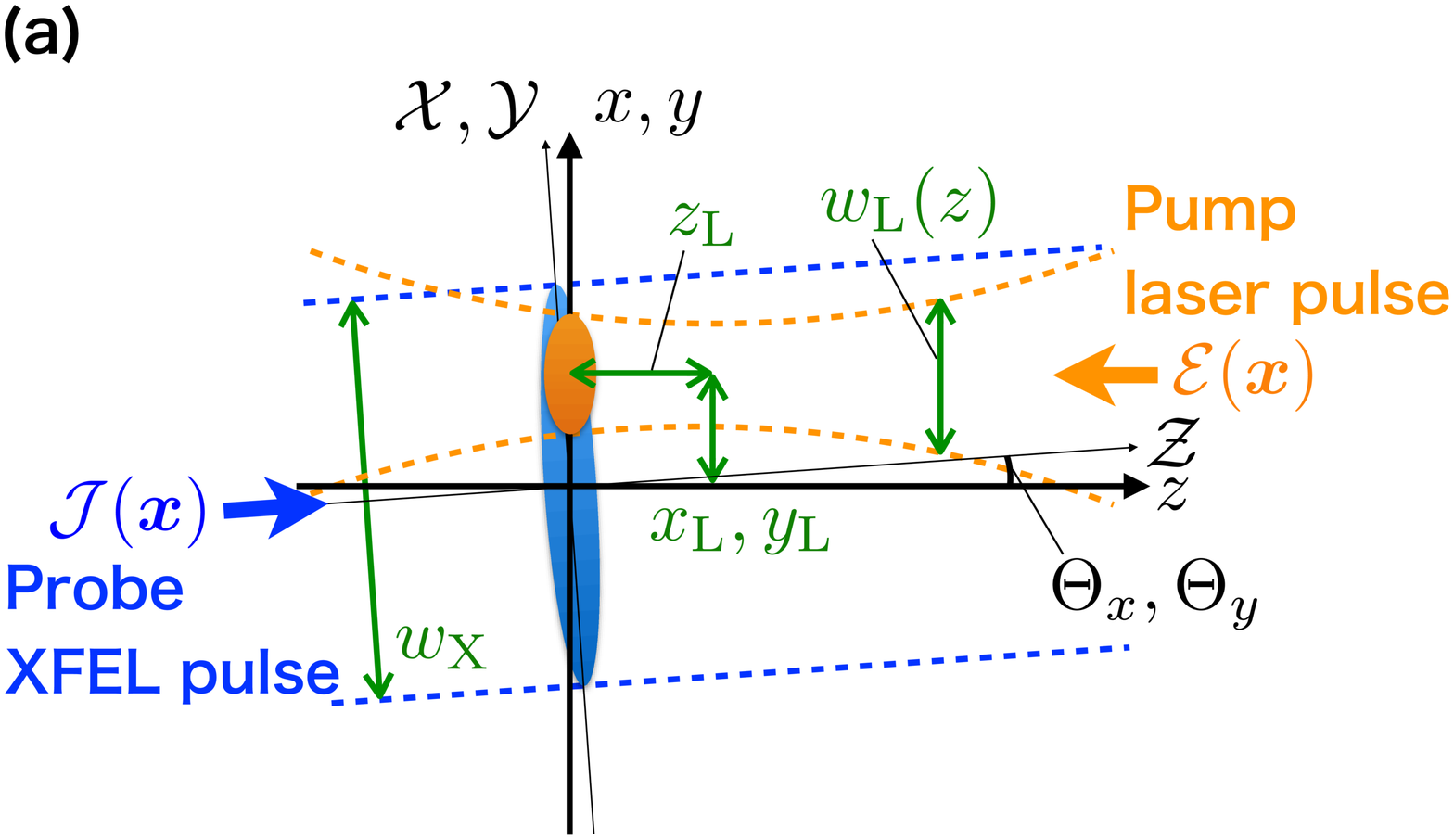}
  \end{minipage}
  \begin{minipage}[b]{0.495\linewidth}
    \includegraphics[width = 1 \linewidth]{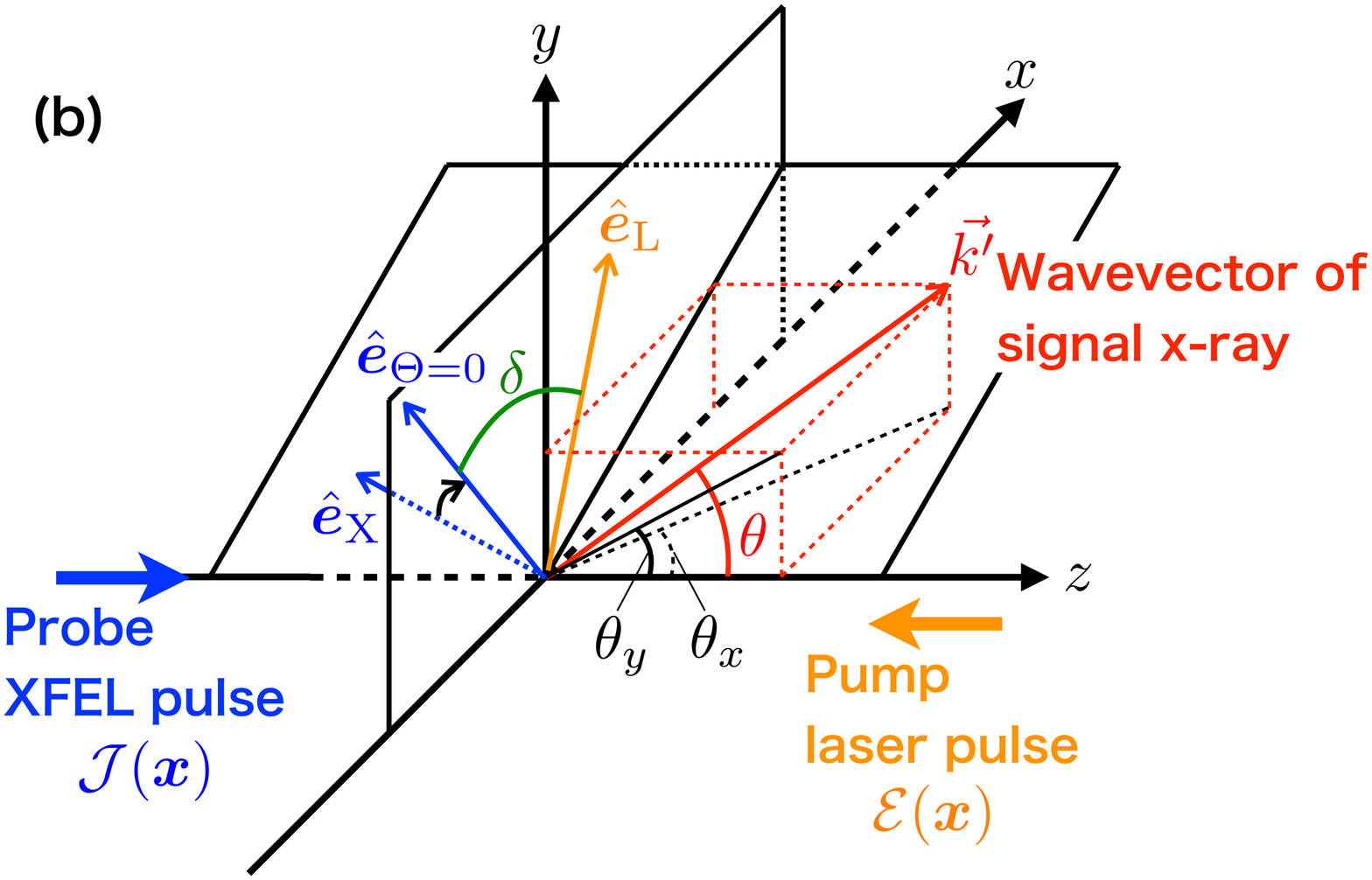}
  \end{minipage}
  \caption{
Schematic drawings of the experimental condition where the XFEL pulse and the laser pulse collide in the head-on geometry.
(a) Magnified view around the collision point from the side.
The beam axis of the XFEL pulse is on the $\mathcal{Z}$-axis, and that of the laser pulse is on the negative direction of the $z$-axis.
  The laser focus has displacements $(x_\text{L}, y_\text{L}, z_\text{L})$ from the origins of each axis.
  The XFEL pulse has incident angles, $\Theta_x,\Theta_y$, from the $z$-axis, and its beam size has been modelled to be constant.
(b) Bird's eye view.
$\hat{\bm{e}}_\text{X}$ ($\hat{\bm{e}}_\text{L}$) is the polarization vector of the XFEL (laser) pulse.
  $\hat{\bm{e}}_{\Theta=0}$, which is on the $x$-$y$ plane, is the polarization vector of the XFEL pulse without the incident angle.
  $\delta$ is an angle between $\hat{\bm{e}}_\text{L}$ and $\hat{\bm{e}}_{\Theta=0}$.
  $\vec{k'}$ is the wavevector of the signal x-ray with a diffraction angle, $\theta$, from the $z$-axis.
  $\theta_x$ ($\theta_y$), is an angle of $\theta$ in the $x$-$z$ ($y$-$z$) plane.
}
  \label{f_CollisionIllustration}
\end{figure*}

A so-called ``pump-probe'' setup illustrated in Fig.\ \ref{f_CollisionIllustration}\ (a) is considered.
The probe XFEL pulse and the pump laser pulse collide in the head-on geometry.
The XFEL pulse has the following field amplitude,
\begin{align}
  {\cal J}(\bm{x}) = & {\cal J}_0 \cos(k(\mathcal{Z}-ct) + \psi_0) \exp\left[-\frac{(\mathcal{Z}/c-t)^2}{(\tau_\text{X}/2)^2}\right] \notag \\
                     & \times \exp\left[-\frac{\mathcal{X}^2 + \mathcal{Y}^2}{w_\text{X}^2}\right],
  \label{e_ProbeXFEL}
\end{align}
with
\begin{align}
  \left(
    \begin{array}{c}
      \mathcal{X} \\
      \mathcal{Y} \\
      \mathcal{Z}
    \end{array}
  \right)
  =
  \cos\Theta
  \left(
    \begin{array}{ccc}
      1 + \frac{\tan^2\Theta_y}{\tan^2\Theta}(\frac{1}{\cos\Theta}-1)         & -\frac{\tan\Theta_x \tan\Theta_y}{\tan^2\Theta}(\frac{1}{\cos\Theta}-1) & -\tan\Theta_x \\
      -\frac{\tan\Theta_x \tan\Theta_y}{\tan^2\Theta}(\frac{1}{\cos\Theta}-1) &  1 + \frac{\tan^2\Theta_y}{\tan^2\Theta}(\frac{1}{\cos\Theta}-1)        & -\tan\Theta_y \\
      \tan\Theta_x                                                            & \tan\Theta_x                                                            & 1
    \end{array}
  \right)
  \left(
    \begin{array}{c}
      x \\
      y \\
      z
    \end{array}
  \right),
\end{align}
where ${\cal J}_0$ is the peak field strength,
$c$ is the speed of light,
$\tau_\text{X}$ is the pulse duration,
$k$ is the wavenumber of the x-ray,
and $\psi_0$ is a constant phase.
The XFEL pulse propagates in the $\mathcal{Z}$-direction and has an incident angle, $\Theta$, against $z$-axis.
$\mathcal{X,Y,Z}$-axis is a rotated $x,y,z$-axis by $\Theta$.
$\Theta_x$ ($\Theta_y$) denotes the incident angle between $\mathcal{Z}$-axis and the $y$-$z$ ($x$-$z$) plane, and satisfies $\frac{1}{\cos^2\Theta} = 1+ \tan^2\Theta_x + \tan^2\Theta_y$.
Equation\ (\ref{e_ProbeXFEL}) represents the XFEL pulse with constant beam sizes, $w_\text{X}$.
This constant assumption on $w_\text{X}$ is reasonable for the calculation of the diffraction
because the beam size is almost constant during the interaction due to the long Rayleigh length of the XFEL pulse.
This XFEL pulse does not have the curvature because of the constant beam size.

The laser pulse is modelled as a pulsed Gaussian beam with a field amplitude,
\begin{align}
  {\cal E} (\bm{x}) = & {\cal E}_0 \cos(\Phi(\bm{x})) \exp\left[-\frac{(z/c+t)^2}{(\tau_\text{L}/2)^2}\right] \notag \\
                      & \times \frac{w_{\text{L}0}}{w_\text{L}(z)} \exp\left[-\frac{(x-x_\text{L})^2+(y-y_\text{L})^2}{w_\text{L}(z)^2}\right] ,
\end{align}
where ${\cal E}_0$ is the peak field strength.
$\Phi(\bm{x})$ is a term which denotes a phase of a Gaussian beam,
and $\tau_\text{L}$ is the pulse duration.
$w_\text{L}(z) = w_\text{L0} \sqrt{1+(\frac{z-z_\text{L}}{z_\text{RL}})^2}$ is the beam size as a function of $z$ with a beam waist, $w_\text{L0}$, and Rayleigh length, $z_\text{RL} = \frac{\pi w_\text{L0}^2}{\lambda}$.
$x_\text{L}$, $y_\text{L}$ and $z_\text{L}$ are displacements from the origin of the $x$-$y$-$z$ coordinate to the laser focus.
Since a photon exchange between the two pulses is well-suppressed, the oscillating term $\cos(\Phi(\bm{x}))$, which depends on the photon energy of the laser, can be averaged over the oscillation of the laser field\cite{c_KarbsteinVD2015, c_KarbsteinVD2016}.
This gives $\cos^2(\Phi(\bm{x})) \rightarrow \frac{1}{2}$.

We calculate the pulse energy of the laser, $W$, to express the peak intensities, ${\cal E}_0^2$ and ${\cal J}_0^2$, with measurable quantities,
\begin{align}
  W = \int \text{d}x\text{d}y\text{d}z \epsilon_0 {\cal E}^2 (\bm{x}) \simeq \epsilon_0 \frac{(2\pi)^{3/2} {\cal E}^2_0 w_\text{L0}^2 c\tau_\text{L}}{32},
\end{align}
where $\epsilon_0$ is the dielectric constant.
From this, the peak intensities can be written as
\begin{align}
  {\cal E}_0^2 \simeq \frac{32W}{(2\pi)^{3/2} \epsilon_0 w_\text{L0}^2 c\tau_\text{L}} \ \ \ \text{and} \ \ \
  {\cal J}_0^2 \simeq \frac{32N \hbar c k}{(2\pi)^{3/2} \epsilon_0 w_\text{X}^2 c\tau_\text{X}},
\end{align}
where $N$ is the number of x rays in the XFEL pulse,
and $\hbar c k$ is photon energy of the x-ray,
and $\hbar$ is the reduced Planck constant.

\subsection{Simplification of signal formula}\label{section2.2}

In the assumed setup, an interaction probability of VD for each x-ray, $P$, is written as\ \cite{c_KarbsteinVD2016},
\begin{align}
  \frac{\text{d}^3P}{\text{d}k'^3} = & \frac{1}{N} \frac{1}{45^2 \pi^3} \frac{\alpha^4 \hbar^5 \epsilon_0^3}{m_e^8 c^9} k' (1+\cos\Theta)^2 (1+\cos\theta)^2 (16+33\sin^2\delta) \notag \\
                                     & \times {{\cal J}_0}^2 {{\cal E}_0}^4 {\left| {\cal M} \right|^2}, \label{e_AngleDistribution_first}
\end{align}
with
\begin{align}
  {\cal M}                         = \int \text{d}^4 {\mathscr{X}} \exp\left(i {\mathscr{K}}'{\mathscr{X}}\right)  \frac{{\cal J} (\bm{x})}{{\cal J}_0} \left(\frac{{\cal E}(\bm{x})}{{\cal E}_0}\right)^2,
\end{align}
where ${\mathscr{X}} = (ct, \bm{x}) = (ct,x,y,z)$ and ${\mathscr K}' = (k', \vec{k'})$ are four-vectors with the metric $(-1,1,1,1)$, and $k' = |\vec{k'}|$.
$\vec{k'}$ is the wavevector of the signal x-ray,
$\theta$ is the diffraction angle of the signal x-ray from the $z$-axis.
$\alpha$ is the fine-structure constant,
$m_e$ is the electron mass.
$\delta$, which is illustrated in Fig.\ \ref{f_CollisionIllustration}\ (b), is the angle between the polarization vector of the laser pulse, $\hat{\bm{e}}_\text{L}$, and a vector, $\hat{\bm{e}}_{\Theta=0}$, where $\hat{\bm{e}}_{\Theta=0}$ is a polarization vector of the XFEL pulse without the incident angle.

To simplify Eq.\ (\ref{e_AngleDistribution_first}), we apply the following approximations and assumptions, which are reasonable in practical experiments:
\begin{itemize}
  \item[(i)] A short-pulse approximation where both pulse durations satisfy $c\tau_\text{L} \ll z_\text{RL}$ and $c\tau_\text{X} \ll z_\text{RL}$.
             This condition is reasonable since a pulse length of typical femtosecond lasers is shorter than the Rayleigh length.
             For example, the pulse length of typical femtosecond lasers satisfies ${c\tau_\text{L} < 30\ \mu}$m (${\tau_\text{L} < 100}$~fs), and $c\tau_\text{X}$ is much shorter than that.
             Whereas $z_\text{RL}$ is $\sim 400$ $\mu$m for $w_\text{L0} = 10\ \mu$m and $\lambda = 800$ nm.
             In this condition, we can approximate $w_\text{L}(z)$ to be constant during the interaction as
\begin{align}
  w_\text{L}(z) \simeq w_\text{L} = w_\text{L0}\sqrt{1+ \left(\frac{z_\text{L}}{z_\text{RL}}\right)^2}. \label{e_ShortPulseApproximation}
\end{align}

\item[(ii)] A small-incident-angle approximation that the incident angle satisfies $\Theta \ll 1$.
  We take only the leading order of $\Theta$.
\item[(iii)] A forward approximation that the diffraction angle satisfies $\theta \ll 1$.
  This condition is reasonable in experiments where an x-ray is used as a probe beam because the diffraction of an x-ray is small due to its short wavelength.
  The approximation enables us to take only the leading order of $\theta$.

\item[(iv)] A cosine term approximation.
  The cosine term in Eq.~(\ref{e_ProbeXFEL}) is expressed as ${\cos(k(\mathcal{Z}-ct) + \psi_0) } = {{\displaystyle \sum_{q=\pm1}} \{ \frac{1}{2} \exp\left[i q \left( k(\mathcal{Z}-ct) + \psi_0 \right) \right] \}}$, and the $q=+1$ term can be ignored because it is substantially suppressed in the $\text{d}k'$ integration \cite{c_KarbsteinVD2016}.
\end{itemize}

By applying (i)--(iv), Eq.\ (\ref{e_AngleDistribution_first}) becomes
\begin{align}
  \frac{\text{d}^3P}{\text{d}k'^3} \simeq & \frac{1}{2\pi} \frac{2^6}{45^2 \pi^2} \frac{\alpha^4 \hbar^6}{m_e^8 c^{10}} (16+33\sin^2\delta) k k' W^2 \notag \\
                                          & \times \frac{1}{w_\text{L}^2 (w_\text{L}^2 + 2 w_\text{X}^2)}
                                            \exp\left[ -\frac{4w^2(x_\text{L}^2 + y_\text{L}^2)}{w_\text{L}^2 w_\text{X}^2} \right] \notag \\
                                          & \times w^2 \exp\left[ -\frac{1}{2} w^2 \left\{(k'\theta_x-k\Theta_x)^2 + (k'\theta_y-k\Theta_y)^2  \right\} \right] \notag \\
                                          & \times \frac{1}{\sqrt{2\pi} \frac{2}{c \tau_\text{X}}}
                                            \exp\left[ - \frac{1}{2} \left(\frac{c\tau_\text{X}}{2} \right)^2 \left( k'-k \right)^2 \right]
                                            \label{e_AngleDistribution_intermediate},
\end{align}
with
\begin{align}
  w^2 = \frac{w_\text{L}^2 w_\text{X}^2}{w_\text{L}^2 + 2 w_\text{X}^2},
\end{align}
where $\theta_x$ $(\theta_y)$ is the diffraction angle of the signal x-ray from the $y$-$z$ ($x$-$z$) plane.
The exponential term in the fourth line in Eq.~(\ref{e_AngleDistribution_intermediate}) is also expressed as
$\exp\left[ - \frac{1}{2} \left(\frac{\tau_\text{X}}{2\hbar} \right)^2 \left(c\hbar k'-c\hbar k \right)^2 \right]$.
The term states that the energy width of the signal x rays is $\frac{2\hbar}{\tau_\text{X}}$.
The energy width, e.g., 0.1 eV given by $\tau_\text{X} = 10$ fs, is narrow compared to an energy width of a typical XFEL $\sim 20$ eV ($1$ standard deviation, 1$\sigma$).
Thus, we can practically neglect the energy spread of the signal x rays and can further apply the following;

\begin{itemize}
\item[(v)] A delta function approximation that the fourth line of Eq.\ (\ref{e_AngleDistribution_intermediate}) is approximated to be $\delta(k'-k)$ in the $\text{d}k'$ integration.
\end{itemize}

Executing the $\text{d}k'$ integration with (v), we obtain the fully simple formula of VD as
\begin{align}
  \frac{\text{d}P}{\text{d}\cos\theta} \simeq & \frac{2^6}{45^2 \pi^2} \frac{\alpha^4 \hbar^6}{m_e^8 c^{10}} (16+33\sin^2\delta) k^2 W^2 \notag \\
                                            &\times \frac{1}{w_\text{L}^2 (w_\text{L}^2 + 2 w_\text{X}^2)} \times \frac{2}{\pi w^2} \int I_\text{IV}(x,y) \text{d}x\text{d}y \notag \\
                                            & \times (kw)^2 \exp\left[ -\frac{1}{2} (kw)^2 \left\{(\theta_x-\Theta_x)^2 + (\theta_y-\Theta_y)^2 \right\} \right],
  \label{e_AngleDistribution_final}
\end{align}
with
\begin{align}
  I_\text{IV}(x,y) = & \exp\left[ -2\frac{x^2 + y^2}{w_\text{x}^2} \right] \notag \\
                     & \times \left( \exp\left[ -2\frac{(x-x_\text{L})^2+(y-y_\text{L})^2}{w_\text{L}^2} \right] \right)^2.
\end{align}
It consists of three parts; the parameter dependency part (the first line), the form part of the two pulses (the second line) and the angular distribution part (the third line).
It provides a useful approximation to understand the phenomenon of VD and to optimize experimental setups since the dependence of the interaction probability on each parameter is clear.

In the second line of Eq.\ (\ref{e_AngleDistribution_final}), $I_\text{IV}(x,y)$ is the product of intensity distributions of the two pulses, $({\cal E}(\bm{x})^2) \times ({\cal J}(\bm{x})^2)^2$, at the collision point.
Exponents agree with the number of photons (one x-ray and two laser photons), which comprises the leading-order Feynman diagram for VD (the QED box diagram).
This overlap of the two pulses can be thought of as an effective interaction volume.
It works as a weight on $P$, showing how good the two pulses overlap in space-time, and gets decreased by large displacements, $(x_\text{L},y_\text{L})$.
$w$ gives the 2$\sigma$ size of this interaction volume.

The third line of Eq.\ (\ref{e_AngleDistribution_final}) shows that the angular distribution has a Gaussian profile with the signal divergence, $\frac{2}{kw}$,
and the center of the distribution is the incident angle.
The uncertainty principle binds a relation between the signal divergence and the size of the interaction volume in the transverse direction; $\Delta p \Delta x = \frac{1}{kw} \hbar k \times \frac{w}{2} = \frac{\hbar}{2}$.

Equation~(\ref{e_AngleDistribution_final}) is also useful to calculate the VB signal, which has the polarization orthogonal to the incident x-ray because the VB signal can be inferred by substituting $(16+33\sin^2\delta) \rightarrow \frac{9}{4}\sin(2\delta)$ \cite{c_KarbsteinVD2016}.

\subsection{Validation of the simple formula}\label{section2.3}

\begin{table}
  \caption{Two sets of parameters used for calculations.
           One is the parameter set of the prototype experiment at SACLA.
           The other is the set in the future experiment planned at SACLA.
           }
  \label{t_ParametersForAngleDistribution}
  \begin{center}
    \begin{tabular}{lrr}\toprule
      \multirow{2}{*}{Parameters}                 &  \multirow{2}{*}{\shortstack{Prototype \\ experiment}} & \multirow{2}{*}{\shortstack{Future \\ experiment}}  \\
                                                  &                  &              \\ \midrule
      XFEL photon energy $\hbar c k$              &          9.8 keV & 9.8 keV      \\
      Laser                                       &     0.6-TW laser & 500-TW laser \\
      Laser pulse energy $W$                      &          0.21 mJ & 12.5 J       \\
      XFEL beam size $w_\text{X}$                 &       $57\ \mu$m & $2\ \mu$m    \\
      Laser beam size $w_\text{L}$                &      $9.8\ \mu$m & $1\ \mu$m    \\
      XFEL pulse duration $\tau_\text{X}$         &           40 fs  & 42 fs        \\
      Laser pulse duration $\tau_\text{L}$        &           17 fs  & 17 fs        \\
      Collision rate                              &           30 Hz  &  1 Hz        \\ \bottomrule
    \end{tabular}
  \end{center}
\end{table}

\begin{table}
  \caption{Comparison of VD signals obtained by Eq.\ (\ref{e_AngleDistribution_final}) and by the previous calculation\cite{c_KarbsteinVD2016}.
           The x-ray flux at SACLA has been assumed to be $3 \times 10^{11}$~photons/pulse.
           }
  \label{t_CalcurationComparison}
  \begin{center}
    \begin{tabular}{llrrr}\toprule
      &                                           &  Prototype experiment                 &  Future experiment   \\ \midrule
      Signal divergence ($2\sigma$)  &   Eq.\ (\ref{e_AngleDistribution_final})  &  $5.84\ \mu$rad                  &  $60.5\ \mu$rad\\
                                     &  previous calculation                     &  $5.84\ \mu$rad                  &  $56.8\ \mu$rad\\ \midrule
      Interaction probability $P$    &   Eq.\ (\ref{e_AngleDistribution_final})  &  $3.9 \times 10^{-26}$           &  $9.4 \times 10^{-12}$\\
                                     &  previous calculation                     &  $2.8 \times 10^{-26}$           &  $8.0 \times 10^{-12}$ \\ \midrule
      Expected signal rate           &   Eq.\ (\ref{e_AngleDistribution_final})  &  $8 \times 10^{-23}$ photons/s   &  0.7 photons/s\\
                                     &  previous calculation                     &  $5 \times 10^{-23}$ photons/s   &  0.6 photons/s \\ \bottomrule
    \end{tabular}
  \end{center}
\end{table}
We validate Eq.~(\ref{e_AngleDistribution_final}) by comparing calculated results with numerical results of the previous calculation \cite{c_KarbsteinVD2016}.
For the calculations, we use two sets of experimental parameters summarized in Tab.~\ref{t_ParametersForAngleDistribution}.
One parameter set is obtained by a prototype experiment, which has been performed in 2017 (Y. Seino et al., manuscript in preparation) at SACLA.
The other is for a future experiment achievable at SACLA.
In the calculations, perfect alignment for both cases is assumed ($\Theta=0$, $x_\text{L} = y_\text{L} = z_\text{L} = 0$), and the calculated results are summarized in Tab.~\ref{t_CalcurationComparison}.
The good agreement of the signal divergence in both parameter sets shows that the angular distribution is well-approximated by the Gaussian profile.
The interaction probability, $P$, also has enough agreements for practical use.
Small gaps are no more than 30\% which is caused due to the change of the function of the laser pulse by the short-pulse approximation (i), as shown in Eq.~(\ref{e_ShortPulseApproximation}).

Expected signal rates are calculated with an angle of acceptance ($18 < \theta_y < 58 \ \mu$rad), which is the same angle with the detection system used in the prototype experiment.
The expected signal rate reaches 0.7~photons/s in the future experiment by assuming $3\times 10^{11}$~photons/pulse of the x-ray flux, which shows that the VD can be observed.

\section{VD formula with XFEL curvature effect}\label{section3}
\begin{figure}
  \centering
  \includegraphics[width = 8.5cm]{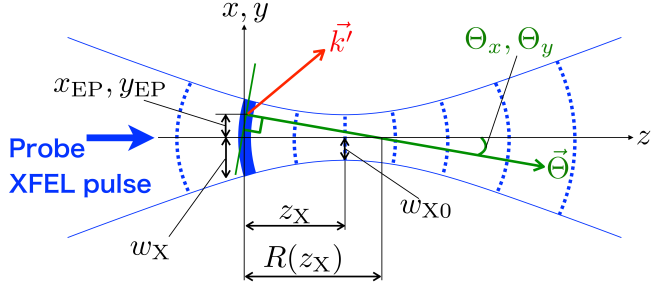}
  \caption{Schematic view of the focused XFEL pulse away from the focus by $z_\text{X}$.
           The wavefront is drawn as a thick line.
           $\Theta_{nx}$ ($\Theta_{ny}$) is an angle between the normal direction of the wavefront, $\vec{\Theta}_n$, and the $y$-$z$ ($x$-$z$) plane.
           These angles effectively give the incident angles to the x-ray.
}
  \label{f_XFELCurvatureIllustration}
\end{figure}

So far, we have considered the simple formula of VD without the curvature.
However, the XFEL pulse has the curvature in practical experiments, and the curvature effectively gives the incident angle to the x-ray.
We consider the XFEL pulse as a Gaussian beam, as shown in Fig.\ \ref{f_XFELCurvatureIllustration}, and the beam size is given as $w_{\rm X} = w_\text{X0} \sqrt{1 + ( \frac{z_\text{X}}{z_{\rm RX}} )^2}$, where $z_\text{X}$ is the distance from the wavefront to the focus along the $z$-axis, $w_\text{X0}$ is a minimum beam size, and $z_\text{RX}$ is Rayleigh length.
The curvature radius of the wavefront is
\begin{align}
  R(z_\text{X}) & = z_\text{X}\left( 1 + \left(\frac{z_\text{RX}}{z_\text{X}} \right)^2 \right).
\end{align}
At a position $(x_\text{EP},y_\text{EP})$ on the wavefront, the normalized normal direction, $\vec{\Theta}_n$, is given as following
\begin{align}
  \vec{\Theta}_n & = \cos\Theta_n \left( \tan \Theta_{nx} ,  \tan \Theta_{ny}  , 1 \right) \notag \\
               & = \left( -\frac{x_\text{EP}}{R(z_\text{X})}, -\frac{y_\text{EP}}{R(z_\text{X})}, \cos\Theta_n \right) \fallingdotseq (\Theta_{nx}, \Theta_{ny}, 1) , \ \ \ \ (\Theta_n \ll 1), \label{e_Theta}
\end{align}
where $\Theta_{nx}$ ($\Theta_{ny}$) is an angle between the normal direction and the $y$-$z$ ($x$-$z$) plane and satisfies $\frac{1}{\cos^2\Theta_n} = 1+ \tan^2\Theta_{nx} + \tan^2\Theta_{ny}$.
These angles are approximately given by the $x,y$ components of the normal direction.

The new VD formula is given by the convolution of the angle of the normal direction and the original angular distribution, Eq.~(\ref{e_AngleDistribution_final}), with the weight, $I_\text{IV}(x_\text{EP},y_\text{EP})$.
The convolution is given by the integrations over $x_\text{EP}$ and $y_\text{EP}$ as
\begin{align}
  \frac{\text{d}P_\text{curve}}{\text{d}x_\text{EP}\text{d}y_\text{EP}\text{d}\cos\theta}
  = & \frac{I_\text{IV}(x_\text{EP},y_\text{EP})}{\int I_\text{IV}(x_\text{EP},y_\text{EP}) \text{d}x_\text{EP}\text{d}y_\text{EP}} \left.\frac{\text{d}P}{\text{d}\cos\theta}\right|_{\Theta_x = \Theta_{nx}, \Theta_y = \Theta_{ny}}, \label{e_CorrectedAngleDistribution}
\end{align}
where VB signal can be also inferred as mentioned in Sec.~\ref{section2.2}.
Due to the convolution, the original divergence, $2/kw$, is smeared by the divergence caused by the curvature at the collision point, $w/R(z_\text{X})$, and the signal divergence becomes their root sum square, $\sqrt{ (2/kw)^2 + (w/R(z_\text{X}))^2}$.
The center of the angular distribution is also shifted to the normal direction at the center of the interaction volume, and the central angles become $(-2w^2 x_\text{L}/w_\text{L}^2 R(z_\text{X}), -2w^2 y_\text{L}/w_\text{L}^2 R(z_\text{X}) )$.

\begin{figure}
  \centering
  \includegraphics[width = 9cm]{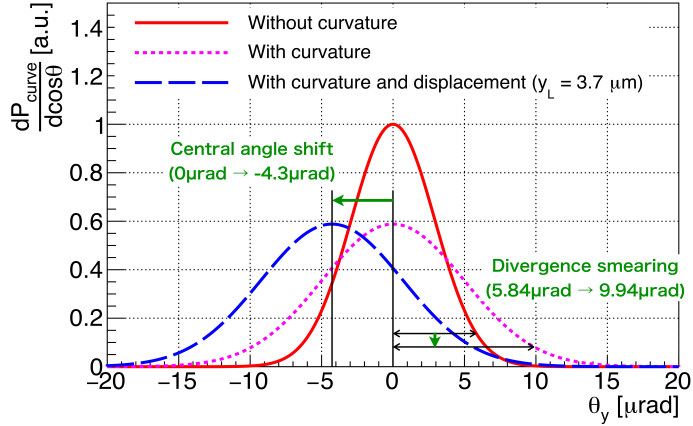}
  \caption{Angular distributions of the signal x rays.
           The solid line: without the curvature effect.
           The dotted line: with the curvature effect.
           The dashed line: with the curvature effect and the displacement of the two pulses.
           The calculations are based on prototype experiment parameters in Tab.~\ref{t_ParametersForAngleDistribution} and Tab.~\ref{t_ParametersForCurvatureCorrection}.
           All distributions are normalized to unity.
}
  \label{f_AngleDistribution_withCurvature}
\end{figure}

\begin{table}[b]
  \caption{Summary of parameters used to estimate the curvature effect for each case in Fig.~\ref{f_AngleDistribution_withCurvature}.
            The forth column corresponds to parameters of the prototype experiment.
  }
  \label{t_ParametersForCurvatureCorrection}
  \begin{center}
    \begin{tabular}{lrrr}\toprule
      \multirow{2}{*}{Parameters}    &  \multirow{2}{*}{\shortstack{Without curvature }} & \multirow{2}{*}{\shortstack{With curvature}} & \multirow{2}{*}{\shortstack{With curvature \\ and displacement}}  \\
                                                   &            &            &               \\ \midrule
      $y_\text{L}$                                 &  0~$\mu$m  &  0~$\mu$m  & 3.7~$\mu$m    \\
      $z_\text{X}$                                 &         -  &  0.85~m    & 0.85~m        \\
      $z_{\rm RX}$                                 &  $\infty$  &  0.09~m    & 0.09~m        \\ \midrule
      signal fraction                              &  \multirow {2}{*}{$2\times10^{-10}$} & \multirow{2}{*}{$1\times10^{-4}$} & \multirow{2}{*}{$3\times10^{-6}$} \\
      (in $18 < \theta_y < 58\ \mu$rad)            &            &            &               \\ \bottomrule
    \end{tabular}
  \end{center}
\end{table}

As an example of the estimation of the curvature effect, we consider the case of the prototype experiment shown in Tab.~\ref{t_ParametersForAngleDistribution}.
In this experiment, the XFEL pulse is focused with an x-ray lens to 57~$\mu$m, and its Rayleigh length, $z_\text{RX}$, is 0.09~m.
The focal point is adjusted at 0.85~m downstream from the collision point ($z_\text{X} = 0.85$~m).
The displacement between the XFEL pulse and the laser pulse is found as $y_\text{L}=3.7$~$\mu$m.
With these parameters, the divergence caused by the curvature is 8.03~$\mu$rad, and it smears the original divergence, 5.84~$\mu$rad, to 9.94~$\mu$rad.
In addition to this, the displacement shifts the central angle of the angular distribution to $-4.3$~$\mu$rad.
The parameters are summarized in Tab.~\ref{t_ParametersForCurvatureCorrection}, and these changes are shown in Fig.~\ref{f_AngleDistribution_withCurvature}.
These changes caused by the curvature effect significantly affect the expected signals.
In the prototype experiment, we set the acceptance of the signal x rays as $18 < \theta _y < 58\ \mu$rad.
As shown in the bottom row of Tab.~\ref{t_ParametersForCurvatureCorrection}, the expected signal fractions are changed more than a few orders of magnitudes.
To avoid these uncertainties, the precise alignments of the beam positions ($x_\text{L},y_\text{L} \ll w$) and the focal position ($z_\text{X} \ll z_\text{RX}$) are required for future high sensitivity experiments.

\section{Conclusion}
We derive the new simple VD and VB formula in the head-on collision geometry of the XFEL pulse and the laser pulse.
The simple formula is the product of terms having physically different origins and helps us to understand the parameter dependencies of the signal easily.
Obtained results by the new formula are compared with those by the previous calculation, and we get good agreements.
We also derive the new VD and VB formula with the wavefront curvature of the XFEL considered, which is first taken into account.
The curvature effect is estimated as the convolution of the XFEL angular distribution and the VD one.
This new formula shows the angular distribution of the signal is broadened and depends strongly on the displacement between the XFEL and the laser foci.

\section*{Acknowledgments}
We are grateful to Felix Karbstein for many fruitful discussions and suggestions related to the numerical analysis.
The research was supported by JSPS KAKENHI Grant Number JP17H02879.
The XFEL experiments were performed at the BL3 of SACLA with the approval of Japan Synchrotron Radiation Research Institute (JASRI) (Proposal No. 2016A8006, 2016B8037).
The synchrotron radiation experiments were performed at BL19LXU in SPring-8 with the approval of RIKEN (Proposal No. 20160031, 20170021).
The research was also supported by a research support program for the graduate student by SACLA.
We would like to thank the beamline staffs of SACLA and SPring-8 for helping our experiment.

\bibliography{Bibliography.bib}

\end{document}